\documentclass{ws-procs975x65}

\begin{document}

\title{Scalar and spinor perturbation to the most generalised Kerr-NUT space-time
}
\author{Banibrata Mukhopadhyay
\footnote{\uppercase{W}ork partially
supported by grant 80750 of the \uppercase{A}cademy of \uppercase{F}inland.}}
\address{Astronomy Division, P.O.Box 3000, University of Oulu,
FIN-90014, Finland; bm@cc.oulu.fi}
\author{Naresh Dadhich}
\address{Inter-University Centre for Astromony and Astrophysics, Post Bag 4,
Ganeshkhind, Pune-411007, India; nkd@iucaa.ernet.in}

\maketitle

\abstracts{ 
We study the scalar and spinor perturbation to Kerr-NUT space-time, that is,
Klein-Gordan and Dirac equation therein. The equations are invariant under
duality transformation between the gravitational electric ($M$) and magnetic ($l$) charge,
radial and angular coordinate, and radial and angular component of the field.
We solve the equations separating into radial and angular parts.
Moreover, if sets of Klein-Gordan and Dirac equation and corresponding solutions are
known for Kerr space-time, under duality transformation, those in dual Kerr space-time
are shown to be achieved. A few examples of solution are shown. 
We comment about the horizon and singularity conditions. 
}


We are familiar with the generalised black hole space-time, Kerr-NUT geometry,
when a rotating black hole contains magnetic charge. This space-time is no longer asymptotically flat
due to presence of magnetic charge.
Very recently, it also has been established that through a set of {\it duality transformation}
over Kerr-NUT metric its form remains invariant\cite{dt}. This {\it duality transformation} is
carried due to the exchange of mass and NUT parameter
in one hand and the radial and angular coordinates on the other hand in a particular fashion.
Moreover, by means of certain transformation properties (over mass and NUT parameter,
and radial and angular coordinates) one can switch over from Kerr to dual Kerr space-time and vice versa.
Dual Kerr is nothing but the exact solution of Einstein equation in vacuum (free of mass parameter).
The present goal is to study the scalar and spinor perturbations to the most
generalised Kerr-NUT space-time by making use of this new duality transformations.

The metric of most generalised Kerr-NUT space-time is
\begin{eqnarray}
ds^2=-\frac{U^2}{\rho^2}\left(dt-Pd\phi\right)^2+\frac{sin^2\theta}{\rho^2}\left[(F+l^2) d\phi
-adt\right]^2 +\frac{\rho^2}{U^2}dr^2+\rho^2d\theta^2,
   \label{met}
   \end{eqnarray}
where
$F=r^2+a^2, U^2=r^2-2Mr+a^2+Q_*^2-l^2$,
$P=asin^2\theta-2lcos\theta, \rho^2=r^2+ \lambda^2,  \lambda = l+acos\theta$.
Here, $l$, $a$ and $Q_*$ are respectively NUT (magnetic charge), angular momentum and electric charge parameters of
the black hole. This metric is invariant under $M\leftrightarrow il, r\leftrightarrow i\lambda$.
Also, $l=0$ gives Kerr space-time; $M=0$ gives dual to Kerr space-time.
$l=0$ and $M\rightarrow il, r\leftrightarrow i\lambda$, brings dual Kerr from Kerr, and
$M=0$ and $l\rightarrow -iM, r\leftrightarrow i\lambda$, brings Kerr from dual Kerr.
According to Newman-Penrose formalism, the set of null tetrads for Kerr-NUT metric can be
written as
\begin{eqnarray}
\nonumber
&&l_{\mu}=\frac{1}{U^2}(U^2, -\rho^2, 0, -U^2 P), \hskip0.3cm
n_{\mu}=\frac{1}{2\rho^2}(U^2, \rho^2, 0, -U^2 P),\\
&&m_{\mu}=\frac{1}{\bar{\rho}\sqrt{2}}(iasin\theta, 0, -\rho^2, -i(F+l^2)sin{\theta}), \hskip0.3cm
{\bar m}_{\mu}=(m_{\mu})^*.
\label{tet}
\end{eqnarray}                                                                                          

When the space-time is perturbed by scalar fields, the Klein-Gordan equation
comes in the picture. Now considering the perturbation as superposition of stationary waves 
with different modes as $\Psi=e^{i(\sigma t+m\phi)}R_0(r)S_0(\theta)$\cite{chanmukh} and 
separating the radial and angular part of Klein-Gordan equation, we get
\begin{eqnarray}
\label{kgeq}
&&\hskip-1cm
\left[\frac{\partial^2}{\partial \theta^2}+cot\theta\frac{\partial}{\partial \theta}-\frac{P^2\sigma^2}
{sin^2\theta}+4l\sigma mcot\theta cosec\theta+\lambda_1^2
+m_p^2(l+acos\theta)^2- \frac{m^2}{sin^2\theta}\right]S_0=0,\\
\nonumber
&&\hskip-1cm
\left[U^2\frac{\partial^2}{\partial r^2}+2(r-M)\frac{\partial}{\partial r}+\frac{\sigma^2}{U^2}(F
+l^2)^2+r^2m_p^2
+\frac{a^2m^2}{U^2}+\frac{2a\sigma m}{U^2}(2l^2-Q_*^2+2Mr)-\lambda_1^2\right]R_0=0.
\end{eqnarray}
The detailed discussions are presented elsewhere\cite{md}. The equation set (\ref{kgeq})
is invariant under $M\leftrightarrow il$,
$r\leftrightarrow i\lambda$  and $R_0\leftrightarrow S_0$. Also $l=0$, $M\rightarrow il$,
$r\leftrightarrow i\lambda$ and $R_0\leftrightarrow S_0$ will bring dual Kerr from Kerr and
vice versa (for $M=0$).

When the space-time is perturbed by spinor fields, the Dirac equation
comes in the picture. Again with the choice of perturbation as stationary wave\cite{chanmukh},
electromagnetic gauge field, $A_\mu\rightarrow \left(-\frac{rQ_{*}}{\rho^2},\,0,\,0,\,
\frac{rQ_* P}{\rho^2}\right)$, and separating the radial and angular part of Dirac equation, we get
\begin{eqnarray}
\nonumber
&&{\it L}_{1/2}S_{1/2}=-({\lambda_2}-m_p(l+acos\theta))S_{-1/2},\hskip0.3cm
{\it L}_{1/2}^{\dagger}S_{-1/2}=({\lambda_2}+m_p(l+acos\theta))S_{1/2},\\
\label{direq}
&&U{\it D}_0R_{-1/2}=(\lambda_2+im_pr)UR_{1/2},\hskip0.3cm
U{\it D}_0^{\dagger}UR_{1/2}=(\lambda_2-im_pr)R_{-1/2},
\end{eqnarray}
where ${\it L}_{n}=\frac{d}{d\theta}+(\sigma P+m)cosec\theta+ncot\theta,
{\it L}_n^{\dagger}=\frac{d}{d\theta}-(\sigma P+m)cosec\theta+ncot\theta$, and
${\it D}_n=\frac{d}{dr}+\frac{i}{U^2}(\sigma (F+l^2)+am+qQ_* r)
+2n\frac{r-M}{U^2}$, ${\it D}_n^{\dagger}=({\it D}_n)^*$. $\lambda_1$ and $\lambda_2$ are
separation constants. The meaning of various
other terms in (\ref{kgeq}) and (\ref{direq}) are discussed elsewhere in detail\cite{md}. 
The set of equations (\ref{direq}) is invariant for $M\leftrightarrow il$,
$r\leftrightarrow i\lambda$ and $iS_{1/2}\leftrightarrow U^{-1/2} R_{-1/2}$ (up and down
spinors transform each other) and vice versa. Like Klein-Gordan case, under the similar
duality transformation Dirac equations also
can be switched over in between Kerr and dual Kerr space-time. 

\begin{figure}[ht]
\includegraphics[height=.41\textheight,width=1.\textwidth]{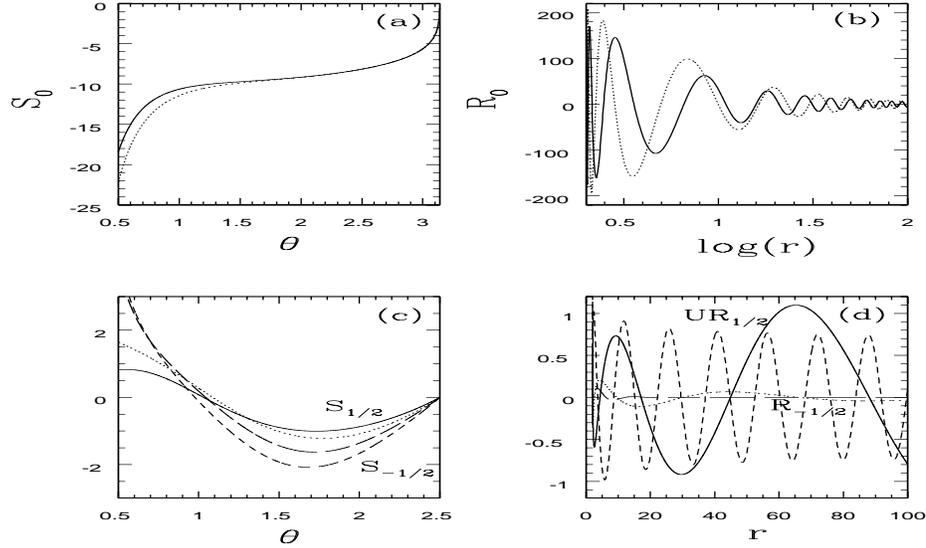}
\caption{Solutions for (a) angular and (b) radial components of scalar perturbation; 
solid and dotted curves are for $m_p=0.4, 0.1$. 
Solutions for (c) angular and (d) radial components of spinor perturbation; 
solid and dotted curves are for spin up with $m_p=0.4, 0.1$ and 
dashed and long-dashed curves are for spin down with $m_p=0.4, 0.1$. $\sigma=0.4$,
$a=0.998$, $l=0.99$, $Q_*=0$, $M=1$.
\label{fig1}}
\end{figure}

Now we like to present a sample set of solutions for scalar and spinor perturbation to the
Kerr-NUT space-time in Fig. \ref{fig1}. Again the detailed solutions for Kerr-NUT, Kerr and
dual Kerr and their comparison are presented elsewhere\cite{md}. Here we choose 
the perturbations of two different rest masses, for a black hole.

Before finishing up our discussions we must comment something about horizon and singularity
(for convenience we choose $Q_*=0$).
The choice of $M=l=0$ and $a\neq 0$ reduces the geometry to flat space-time in vacuum
without horizon. 
On the other hand, for only $l=0$, Kerr (with $M^2 \ge a^2$), space-time has a singularity when 
$\rho^2=r^2+\lambda^2=0$. For $r=0,\theta=\pi/2$, it has a ring singularity where curvatures
diverge. Also horizon forms at $r=r_+,r_-$. For $a>l\neq 0$, regular Kerr-NUT,
singularity at $r=0$ only can be formed when $cos\theta=\pm l/a$. Therefore, for $l>a$ singularity
does not exist for $r\ge 0$, when inner horizon, $r_-<0$. If $l=a$, singularity arises
for $\theta=0,\pi$ and horizons are at, $r_+=2M$, $r_-=0$. For dual Kerr, $M=0$, $r_\pm=\pm\sqrt{l^2-a^2}$.
Therefore for dual Kerr the picture of canonical black hole of horizon covering singularity only can be 
achieved if $l=a$, with the coexistence of real horizon and singularity at $r=0$ and $\theta=0,\pi$.

\end{document}